\documentclass[prl,floatfix,twocolumn,showpacs,preprintnumbers,longbibliography]{revtex4-1} 

\usepackage[utf8]{inputenc} 
\usepackage{bibunits}
\usepackage{amssymb,amsfonts,amsmath,mathrsfs,bbold}

\usepackage[normalem]{ulem}

\usepackage[usenames,dvipsnames]{xcolor}
\usepackage[colorlinks=true,citecolor=Blue,urlcolor=Blue,linkcolor=Blue]{hyperref}

\usepackage[all]{hypcap}
\usepackage{multirow}
\usepackage{graphicx}
\usepackage{physics}

\newcommand{\diag}{\operatorname{diag}}

\begin{document}

\title{Diagnosing non-Hermitian Many-Body Localization and Quantum Chaos\\ via  Singular Value Decomposition}

\author{Federico Roccati}
\thanks{These authors contributed equally to the work.}
\author{Federico Balducci}
\thanks{These authors contributed equally to the work.}
\author{Ruth Shir}
\thanks{These authors contributed equally to the work.}
\author{Aur\'elia Chenu}

\affiliation{Department of Physics and Materials Science, University of Luxembourg, L-1511 Luxembourg}
    
\begin{abstract}
    Strong local disorder in interacting quantum spin chains can turn delocalized eigenmodes into localized eigenstates, giving rise to many-body localized phases. This is accompanied by distinct spectral statistics: chaotic for the delocalized phase and integrable for the localized phase. In isolated systems, localization and chaos are defined through a web of relations among eigenvalues, eigenvectors, and real-time dynamics. These may change as the system is made open. We ask whether random dissipation (without random disorder) can induce chaotic or localized behavior in an otherwise integrable system. The dissipation is described using non-Hermitian Hamiltonians, which can effectively be obtained from Markovian dynamics conditioned on null measurement. In this non-Hermitian setting, we argue in favor of the use of the singular value decomposition. We complement the singular value statistics with different diagnostic tools, namely, the singular form factor, and the inverse participation ratio and entanglement entropy of  singular vectors. We thus identify a crossover  of the singular values, from chaotic to integrable spectral features, and of the singular vectors from delocalization to localization. Our method is illustrated in an XXZ Hamiltonian with random local dissipation.
\end{abstract}

\maketitle


Since the early days of quantum mechanics, understanding the dynamics of many-body quantum systems continues to be a hard challenge. One of the chief questions on the behavior of interacting quantum systems concerns the presence of quantum chaos~\cite{Bohigas1984Characterization,Deutsch1991Quantum,Srednicki1994Chaos,Haake2010Quantum}. Additionally, the extension of quantum localization~\cite{Anderson1958Absence} to interacting systems~\cite{Gornyi2005Interacting,Basko2006Metal} has led to postulating the existence of a robust, non-ergodic phase of matter known as \emph{many-body localization} (MBL). The competition between localized and chaotic quantum dynamics has been studied extensively in spin Hamiltonians~\cite{Oganesyan2007Localization, Znidaric2008Many, Pal2010Many, Bardarson2012Unbounded,SierantPRL2020}, with relevant implications for applications, including quantum annealing~\cite{Altshuler2009Adiabatic, *Altshuler2010Anderson,Laumann2015Quantum}, as well as fundamental questions, such as the lack of thermalization~\cite{Serbyn2013Local,Ros2015Integrals,Imbrie2016Diagonalization,Imbrie2016Local,DAlessio2016Quantum,Alet2018Many,Abanin2019Colloquium}. As a result, localization has become central for understanding complex quantum dynamics, with connections to quantum simulation experiments~\cite{Schreiber2015Observation, Smith2016Many, Lukin2019Probing}, topological phases of matter~\cite{Huse2013Localization, Chandran2014Many, Parameswaran2018Many}, Floquet time crystals~\cite{Khemani2016Phase}, and many others.

The peculiarities in the dynamics of many-body quantum systems are not limited to the ideal situation where the system is isolated from the environment and the dynamics is unitary~\cite{Mezard1987Spin,Biella2021Many,Turkeshi2021Measurement,Youenn2023Volume,TurkeshiPRB2023}. In the last few years, 
the conventional understanding of renormalization group approaches---by which the coupling to a thermal bath would render quantum fluctuations irrelevant~\cite{Sachdev2011Quantum}---has been shown to be incomplete. Indeed, evidence is being accumulated that open quantum systems may host unusual phases that would exist neither in a quantum unitary setting nor at equilibrium~\cite{Li2018Quantum,Skinner2019Measurement,Matsumoto2022Embedding,KawabataPRB2022}. 

One of the many intriguing features of open quantum system dynamics is the phenomenon of dissipative localization~\cite{Hatano1996Localization,*Hatano1997Vortex,*Hatano1998NonHermitian,Goldsheid1998Distribution,Kawabata2021Nonunitary,YamamotoPRB2023} and dissipative quantum chaos~\cite{AkemannPRL2019,AkemannPRE2022,Cornelius2022Spectral,Kawabata2023Symmetry}. For the unitary counterpart, both localization and chaos are defined through a web of relations among eigenvalues, eigenvectors, and real-time dynamics~\cite{Haake2010Quantum,Imbrie2016Local,Alet2018Many,Abanin2019Colloquium,PauschPRL2021}. As the constraints set by unitarity are lifted, it is natural to expect that such relations may change in nature. In particular, open dynamics conditioned to no jumps, described by effective \textit{non-Hermitian} (NH) Hamiltonians, are being thoroughly investigated. While NH localization is fairly well understood in the single-particle case, which admits exact solutions~\cite{DeTomasi2022NonHermitian,*DeTomasi2023NonHermiticity} and a clear renormalization-group treatment in one dimension~\cite{Kawabata2021Nonunitary}, its many-body version has been the object of several numerical studies, suggesting the presence of a stable, localized phase~\cite{Hamazaki2019NonHermitian,Zhai2020Many,Suthar2022NonHermitian,Ghosh2022Spectral,hamazaki2022lindbladian,detomasi2023stable}. These studies mostly relied on the \textit{eigendecomposition} of large NH matrices. However, because of non-Hermiticity, the  indicators of Hermitian localization had to be generalized, causing some ambiguity given the complex nature of eigenvalues~\cite{HamazakiPRRes2020,Sa2020Complex} and the non-orthonormality of right and left eigenvectors~\cite{lu2023unconventional}. Recent works put forward the idea that using the \textit{singular value decomposition}, one can circumvent certain problems set by the eigendecomposition of NH Hamiltonians~\cite{Brody2014Biorthogonal,HerviouPRA2019,PorrasPRL2019,BrunelliSciPost2023}, since the left and right singular vectors are always orthonormal and the singular values are always real. This approach was benchmarked against standard random matrix ensembles~\cite{Kawabata2023Singular} and has not yet been used to study the localization transition of many-body open quantum systems in finite dimensions.

In this work, we fill the gap by studying NH many-body localization via the singular value decomposition. Our objective is twofold. First, we show that the singular value decomposition clearly distinguishes between the chaotic and localized regimes in NH models, providing cleaner and more robust numerical indicators than those obtained from the spectrum. Second, we show that random local dissipation in an otherwise integrable XXZ Hamiltonian induces quantum chaos for small dissipation strength, followed by localization for large dissipation. These crossovers are similar to the ones caused by a purely Hermitian disorder, so our results provide one more point of contact between Hermitian and NH MBL.


\textit{Model.---}The tools we introduce to diagnose NH quantum chaos and MBL are illustrated in a model made 
of an integrable interacting Hermitian term, and a disordered non-Hermitian contribution describing random site-dependent losses, namely,
\begin{equation}\label{eq:model}
    \hat H  = \hat H_\text{XXZ} - i \hat \Gamma/2\,,
\end{equation}
with
\begin{subequations}
    \begin{align}
        \label{eq:XXZ}
        \hat H_\text{XXZ} & = J \sum_{i=1}^N \left( \hat S_i^x \hat S_{i+1}^x + \hat S_i^y \hat S_{i+1}^y + \Delta \hat S_i^z \hat S_{i+1}^z  \right), \\
        \label{eq:loss}
        \hat \Gamma & = \sum_{i=1}^N \gamma_i \left( \hat S^z_i +1/2\right).
    \end{align}
\end{subequations}
Above, $\hat{S}_i^{x,y,z} = \mathbb{1}_2^{\otimes (i-1)} \otimes \frac{1}{2}\hat \sigma^{x,y,z} \otimes \mathbb{1}_2^{\otimes (N-i)}$ are spin-1/2 operators acting on  site $i$ ($\hat{\sigma}^{x,y,z}$ are the Pauli matrices). The coefficient $J$ is set to unity, fixing the energy scale, and we  take $\Delta = 1$. The rates $\gamma_i$ are independently sampled from a uniform distribution over the interval $[0,\gamma]$. We assume periodic boundary conditions; this does not influence the system's behavior since the hoppings are symmetric (contrary to the Hatano-Nelson model~\cite{Hamazaki2019NonHermitian}, which we also analyze in the Appendix~\cite{SupplMat}). As the magnetization is conserved, we choose to work in the zero magnetization sector, of dimension $D =\binom{N}{N/2}$.

The XXZ Hamiltonian~\eqref{eq:XXZ} is an integrable many-body system that has been extensively studied when complemented with random local magnetic fields $\sum_i h_i\hat S^z_i$, where the $h_i$'s are random variables, \textit{e.g.}, uniformly distributed over $[-h,h]$~\cite{Znidaric2008Many,Luitz2015Many}. As such, it has been used to probe a transition between chaos and integrability,  occurring as a function of the disorder strength~\cite{bertrand2016anomalous,Sierant2019Level}. 
The XXZ chain with weak disorder exhibits chaotic spectral properties, described by the \textit{Gaussian orthogonal ensemble} (GOE), and delocalized eigenstates, while in the presence of a strong disorder, it shows integrable spectral properties and localized eigenstates, at least for the system sizes accessible by numerics. 
By contrast, the existence of a finite-disorder MBL phase with local integrals of motion~\cite{Serbyn2013Local,Ros2015Integrals,Imbrie2016Diagonalization} in the thermodynamic limit is still debated~\cite{Suntajs2020Quantum,Panda2020Can,Abanin2021Distinguishing,Sierant2022Challenges,Morningstar2022Avalanches,Crowley2022Constructive,Sierant2024Many}.

The NH Hamiltonian $\hat H$ we consider here can be obtained from the full Lindblad master equation with coherent dynamics driven by $\hat H_\text{XXZ}$, and dissipative dynamics dictated by the quantum jump operators $\sqrt{\gamma_i}\hat S^-_i$, where  $\hat S_i^\pm = \hat S_i^x \pm i\hat S_i^y$. The Lindblad equation can be regarded as the unconditional evolution of the system, that is, averaged over a large number of trajectories~\cite{ZwanzigPhysRev1961,Brun2002Simple,MingantiPRA2019,Ciccarello2022Quantum}. Focusing only on the no-jump trajectories only (null-measurement condition), the open system dynamics is described through the effective NH Hamiltonian $\hat H$~\cite{SupplMat}. In turn, this physical origin of $\hat H$ leads to non-negative jump rates $\gamma_i \geq 0$~\cite{Breuer2007Theory}.

Our investigation uses the Hamiltonian \eqref{eq:model} as a toy model, but our methods are not model-dependent. Similar results were found for another commonly considered NH model, the interacting Hatano-Nelson model~\cite{Hamazaki2019NonHermitian}, as detailed in~\cite{SupplMat}.


\textit{Eigenvalue vs singular value decomposition.---}Hermitian Hamiltonians have real eigenvalues and orthonormal eigenstates, whose physical meanings are the possible energy measurement outcomes and the corresponding quantum states, respectively. For this reason, the \textit{eigendecomposition} (ED) of a Hermitian Hamiltonian, $\hat H=\sum_n E_n\dyad{w_n}$, has a fundamental role in quantum mechanics. Importantly, the ED is realized with a single unitary operator~\cite{SupplMat}.

In turn, NH Hamiltonians  have generally complex eigenvalues, and non-orthogonal eigenvectors. 
For a diagonalizable NH Hamiltonian $\hat H$, the ED can be generalized with the tools of \textit{biorthogonal quantum mechanics}~\cite{Brody2014Biorthogonal,Miri2019Exceptional}. 
This approach has been  used to (attempt to) generalize many known results from the Hermitian setting~\cite{KunstPRL2018}: using the eigenvectors of $\hat H^\dagger$, which are orthogonal to those of $\hat H$, it is possible to resolve the identity and diagonalize the Hamiltonian using two different non-unitary  operators, namely, $\hat H=\sum_n E_n \dyad{R_n}{L_n}$. 

Regardless, a complex spectrum and the use of both right and left eigenvectors (those of $\hat H$ and $\hat H^\dagger$, respectively) pose a challenge to the generalization of certain quantities that are well defined in the Hermitian case, as they rely on a real spectrum and orthonormal states. For instance, we show  that complex spectral gap ratios~\cite{Sa2020Complex} provide a less clear distinction between chaotic or integrable dissipative models~\cite{SupplMat}. Also, the ambiguity in the definition of the density matrix corresponding to a pure state $\ket{R_n}$ ($\dyad{R_n}{R_n}$ or $\dyad{R_n}{L_n}$) is reflected in the different choices made in the definitions of topological invariants, for which both right and left eigenvectors are typically used~\cite{YaoPRL2018}; in the definition of the entanglement entropy, commonly used in NH MBL works~\cite{Hamazaki2019NonHermitian} and based on only the right (or left) eigenvectors; and even in the definition of the inverse participation ratio, for which all combinations have been considered~\cite{DeTomasi2022NonHermitian} without reaching a consensus~\cite{SupplMat}.

Recently, the attention on the \textit{singular value decomposition} (SVD)~\cite{Trefethen1997Numerical}, which can be regarded as a generalized version of the ED, has been growing. Note that, for a (non-)Hermitian Hamiltonian $\hat H$, its ED and the SVD are (not) related~\cite{SupplMat}. 

The SVD, namely $\hat H=\sum_n \sigma_n \dyad{u_n}{v_n}$, has the advantage, compared to the ED,  of providing real (non-negative) singular values $\{\sigma_n\}$ and two sets of (independently) orthonormal singular vectors, $\{\ket{u_n}\}$ and $\{\ket{v_n}\}$~\cite{SupplMat}. Although the biorthogonal left and right eigenvectors of a NH Hamiltonian can be made \textit{biorthonormal}, $\langle{L_n}\vert{R_m}\rangle=\delta_{nm}$, it is not possible to normalize both simultaneously. In contrast, the left and right singular vectors are automatically normalized, therefore corresponding to physical states. This represents a strong motivation to use the SVD for NH Hamiltonians to generalize and study well-established Hermitian phenomena. Indeed, the SVD has been shown to be instrumental in describing the bulk-boundary correspondence in NH topological models~\cite{HerviouPRA2019,PorrasPRL2019,RamosPRA2021,BrunelliSciPost2023,GomezLeon2023drivendissipative}, and in studying the statistics of NH random matrices as a measure of dissipative quantum chaos~\cite{Kawabata2023Singular}. 

For these reasons, we use the SVD to study the model~\eqref{eq:model}. Its Hermitian version exhibits MBL and chaotic behavior for strong and weak disorder, respectively. Here, we investigate whether random non-Hermiticity, which physically corresponds to random losses [see Eq.~\eqref{eq:loss}], can induce a chaotic to integrable crossover. We do so by using singular-value statistics~\cite{Kawabata2023Singular} and the \textit{singular form factor}, a measure of correlations we define below. Furthermore, we study the localization transition of the singular vectors. Our results further motivate the use of the SVD as a sensitive tool to generalize Hermitian phenomena, such as MBL, to the NH setting.

\textit{Dissipative quantum chaos: the singular form factor.---}One of the defining features of quantum chaos is the level spacing distribution, which is well-known in random Hermitian Hamiltonians taken from Gaussian ensembles  (orthogonal, unitary, and symplectic) and Hamiltonians whose eigenvalues are not correlated (Poisson ensemble)~\cite{Mehta2004Random,Haake2010Quantum}. Indeed, the spectrum of a chaotic Hamiltonian is conjectured to have a level spacing distribution that follows random matrix behavior~\cite{Bohigas1984Characterization}, while the spectrum of an integrable Hamiltonian is uncorrelated, and its level spacing is expected to follow an exponential (but usually referred to as Poisson) distribution~\cite{Berry1977Level}. However, to be able to make such statements for a specific system,  the spectrum must be unfolded before one computes the level spacing distribution to remove the global energy dependence of the eigenvalue density~\cite{Haake2010Quantum}. Two alternative measures to extract information about the onset of chaos, while circumventing the unfolding procedure, are the \textit{spectral form factor} (SFF)~\cite{leviandier1986fourier, wilkie1991time, Alhassid1992spectral, Ma1995CoreHole, Cotler2017Black, delCampo2017SFF, Chenu2018Quantum, Xu2021Thermofield, Chenu2023knSFF} and the spectral ratio statistics~\cite{Oganesyan2007Localization,Atas2013Distribution}. 

\begin{figure}
    \centering
    \includegraphics[width=\columnwidth]{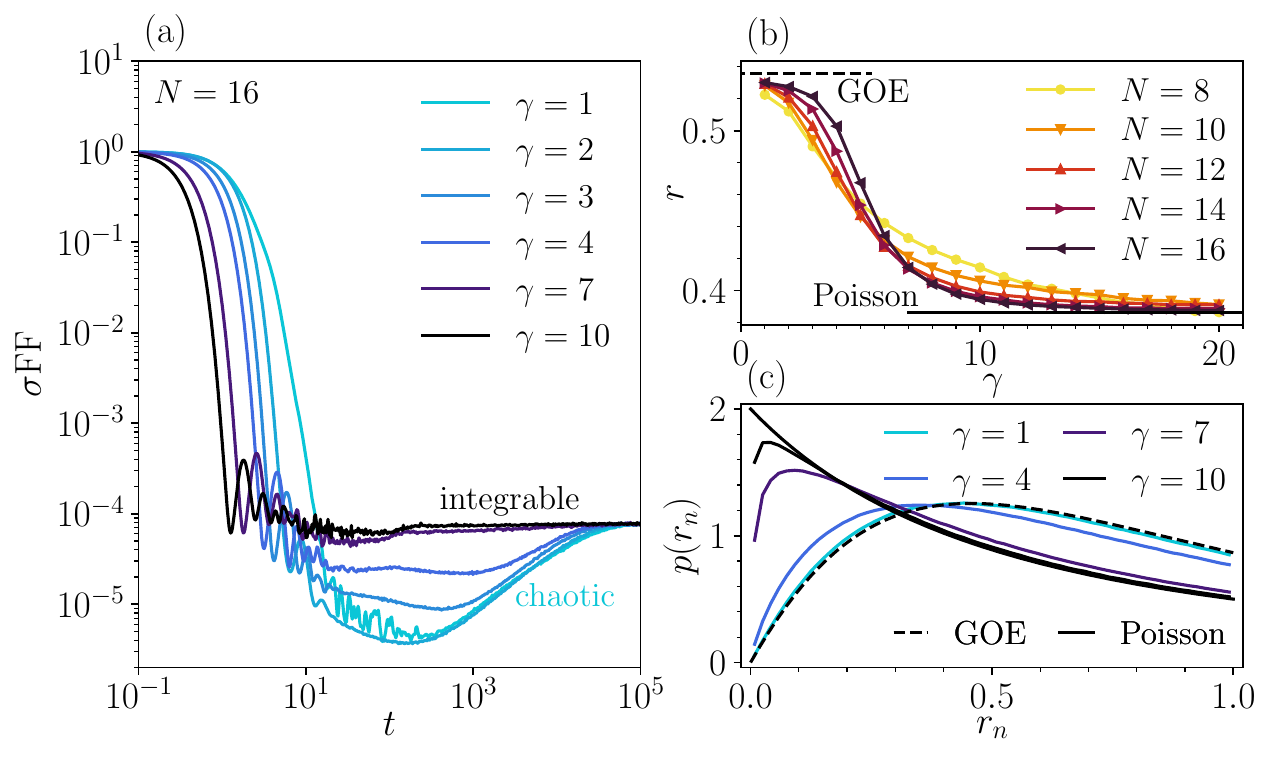}
    \caption{\textit{Dissipative quantum chaos through the SVD.} (a) Singular form factor ($\sigma$FF), Eq.~\eqref{eq:sigmaFF}, for the XXZ model with random losses \eqref{eq:model}. For small dissipation strength a ramp is present, signaling repulsion between singular values, while for stronger dissipation, the correlation hole disappears, leaving only a plateau, as in the integrable case (black). (b) Ratio statistics for the singular values, computed using a portion of the smallest singular values. As the dissipation is made stronger, the average of the ratio distribution signals a crossover from chaos (GOE value) to integrability (Poisson value). Error bars are smaller than symbols. (c) The crossover is also apparent in the full probability distribution of the ratios (shown for $N=16$). All the data are averaged over at least 7000 disorder realizations.}
    \label{fig:values}
\end{figure}

For a standard Hermitian Hamiltonian $\hat H$ with eigendecomposition $(\hat{H}-E_n) \ket{w_n}=0$, we recall that the spectral form factor is defined as $\text{SFF}(t) = | \sum_n e^{-iE_nt}/D |^2$, where $D$ is the Hilbert space dimension. One may also express it as $\text{SFF}(t) = |\!\bra{\psi} e^{-i\hat H t}  \ket{\psi}\!|^2$, that is, as the return probability of the infinite-temperature \textit{coherent Gibbs state}, $\ket{\psi}=\sum_n \ket{w_n}/\sqrt{D}$. This form is particularly handy and has been used to generalize the SFF to dissipative and non-Hermitian dynamics~\cite{Cornelius2022Spectral,Matsoukas2023NonHermitian}. The SFF is a time-dependent quantity with distinct features at different time scales. In both integrable and chaotic systems, the ensemble-averaged SFF decays at early times and saturates to a plateau of value $1/D$ at very late times~\footnote{Here we are assuming that the energies have no special relations among themselves, as is generically expected in many-body, interacting quantum systems.}. Its behavior in-between differentiates integrable systems, which go directly from decay to plateau, from chaotic systems, in which the SFF exhibits a correlation hole followed by a  linear growth before the plateau. This additional ``ramp'' stems from correlations between eigenvalues; it is visible whether or not the spectrum is unfolded~\cite{Cotler2017Black, santos2020speck}. 

We generalize this return probability to a NH Hamiltonian $\hat H$ by introducing the \textit{singular form factor} ($\sigma$FF): 
\begin{equation}
    \label{eq:sigmaFF}
    \!\!\! \sigma\text{FF}(t) = \Bigg| \frac{1}{D}\sum_n e^{-i\sigma_nt} \Bigg|^2 = |\!\bra{\psi_\mathrm{R}} e^{-i\sqrt{\hat H^\dagger \hat H} t} \ket{\psi_\mathrm{R}}\!|^2\,.
\end{equation}
This extends the SFF to the NH case via the SVD: $\sigma_n$ are the singular values of $\hat H$ and $\ket{\psi_\mathrm{R}}=\sum_n \ket{v_n}/\sqrt{D}$ is the \textit{right} infinite temperature coherent Gibbs state, built from its right singular vectors  $\ket{v_n}$~\cite{SupplMat}. Note that Eq.~\eqref{eq:sigmaFF} can also be written in terms of the left singular vectors $\ket{u_n}$, replacing $\ket{\psi_\mathrm{R}}$ with $\ket{\psi_\mathrm{L}}\!=\!\sum_n \ket{u_n}/\sqrt{D}$ and $\hat H$ with $\hat H^\dagger$.

We argue that the $\sigma$FF is a good indicator of quantum chaos in a NH setting, being able to detect the presence of correlations among singular values. 
Figure~\ref{fig:values}a  shows the $\sigma$FF for various disorder strengths in the considered model, Eq.~\eqref{eq:model}. The $\sigma$FF exhibits a correlation hole before the plateau for small disorder. This correlation hole closes as the disorder strength gets larger, indicating the loss of correlations between the singular values.

In parallel to the form factor, the distribution of the spectral ratios $r_n=\min(s_{n+1},s_n)/\max(s_{n+1},s_n)$, where $s_n=E_{n+1}-E_n$ are the level spacings between ordered eigenvalues, is also used as a spectral probe of chaos vs integrability~\cite{Oganesyan2007Localization,Atas2013Distribution}. Because it involves ratios of level spacings, the density of states cancels out, removing the need for unfolding to compare systems with different global densities. Distributions of $r_n$ are known for the Gaussian and Poisson ensemble~\cite{Atas2013Distribution}. In our case, the relevant ensembles are the GOE for low disorder and the Poissonian ensemble for high disorder. The probability density distributions $p(r_n)$ and their average values $r$ can be found in Ref.~\cite{Atas2013Distribution}.

The statistics of the singular values can also be studied via the spectral ratios defined above, replacing $E_n$ with $\sigma_n$ in the definition of the level spacing $s_n$~\cite{Kawabata2023Singular}. This idea was recently used to classify the singular-value statistics of NH random matrices~\cite{Kawabata2023Singular}, and we extend it to study the chaotic to integrable crossover. In this respect, Figs.~\ref{fig:values}b--c clearly display a crossover from GOE to Poisson statistics as the dissipation strength $\gamma$ is ramped up. In particular, Fig.~\ref{fig:values}b suggests the presence of a finite-size crossover around $\gamma_c/J \gtrsim 9$. In~\cite{SupplMat}, we further show that, in the NH case, the results for the generalizations of the ratio statistics to complex eigenvalues~\cite{Sa2020Complex} are rather vague compared with the results for the singular values. The $\sigma$FF we introduce, together with the singular value spacing statistics, show how the SVD is an appropriate tool to detect a chaotic to integrable crossover in NH quantum systems. Furthermore, our results point to the occurrence of a localized regime, as detailed below.


\textit{Dissipative localization of singular vectors.---}The singular-value indicators presented in Fig.~\ref{fig:values} support the presence of a localized regime in the XXZ model with random dissipation, Eq.~\eqref{eq:model}. The localization induced by disorder, however, is better understood from a real-space perspective, as the name itself suggests. It is thus interesting to see whether the singular vectors of NH models display the same signatures of localization as their Hermitian counterparts (or even as NH eigenvectors). 

\begin{figure}
    \centering
    \includegraphics[width=\columnwidth]{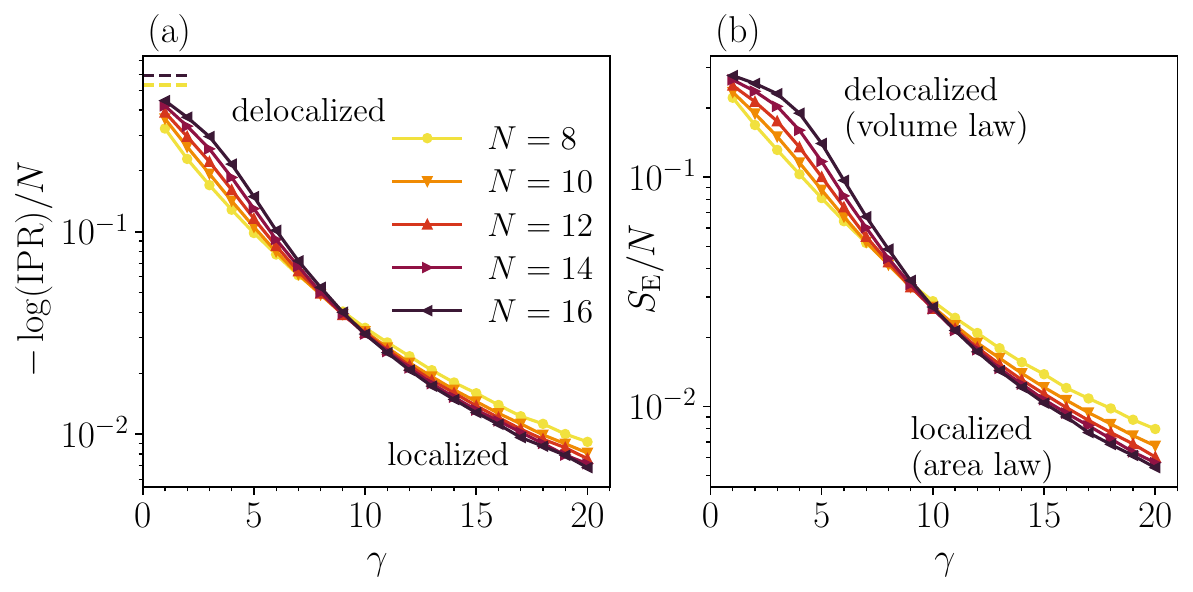}
    \caption{\textit{Dissipative localization of singular vectors.} (a) The inverse participation ratio (IPR) quantifies how much a state is localized in the many-body Fock space, passing from IPR $=O(1/D)$ (value illustrated with the dashed lines; delocalized regime) for weak dissipation to IPR $=O(1)$ (localized) for strong dissipation. Our data shows a crossover between the two regimes as the disorder in the dissipator is ramped up. (b) Entanglement entropy $S_\mathrm{E}$ across a bipartition of the chain in two halves. For small dissipation $\gamma$, $S_\mathrm{E}$ increases at least with the system size $N$ (volume law, delocalized states), while at larger $\gamma$ the entanglement entropy decreases with system size when divided by $N$, thus suggesting an area law of entanglement (localized states). The data are averaged over at least 7000 disorder realizations, and error bars are smaller than symbols.}
    \label{fig:vectors_scaled}
\end{figure}

For a single particle, the eigenstates of a Hermitian and localized Hamiltonian have a well-understood real-space structure. Each eigenstate $\ket{w_n}$ is concentrated around its localization center $\mathbf{x}_n$, and its decaying profile is characterized by a localization length $\xi$, namely $\braket{\mathbf{x}}{w_n} \sim e^{-|\mathbf{x}-\mathbf{x}_n|/\xi}$. A similar situation takes place in single-particle, NH, localized Hamiltonians, in which the disorder-induced localization competes with the localization yielded by the non-Hermitian skin effect~\cite{Hatano1996Localization,Hatano1997Vortex,Hatano1998NonHermitian}.  

In the many-body case, the situation is more complicated. Even in the Hermitian setup, there seems to be no simple localization in Hilbert space~\cite{DeLuca2013Ergodicity}. Rather, the eigenstates of MBL Hamiltonians are believed to be eigenstates of \emph{local integrals of motion}~\cite{Serbyn2013Local,Ros2015Integrals,Imbrie2016Diagonalization,Imbrie2016Local} as well and to obey the area law of entanglement~\cite{Abanin2019Colloquium}.

Previous works studied some aspects of NH, localized, many-body eigenvectors, \textit{e.g.}, identifying a crossover from volume to area law for the entanglement entropy~\cite{Hamazaki2019NonHermitian}. Here, we perform a fundamentally different analysis, studying the localization of singular vectors of NH models. These vectors, having all the properties of physical states, do not suffer from the ambiguities of right and left eigenvectors~\cite{SupplMat}. We use the SVD to study localization, and show it can discern between the localized and ergodic regimes. As such, it extends the use of SVD beyond diagnosing dissipative chaos~\cite{Kawabata2023Singular}.

For simplicity, we use two commonly considered indicators: the \textit{inverse participation ratio} (IPR) and the entanglement entropy across a bipartition. Our analysis is based on the right singular vectors; using the left ones yields similar results.

The IPR of singular vectors is defined as the ensemble average of $\sum_{k=1}^D |\!\braket{e_k}{v_n}\!|^4 /D$, where $\{\ket{e_k}\}$ is the computational basis and $\ket{v_n}$ are the (right) singular vectors.
It is expected that $\mathrm{IPR} = O(1/D)$ for delocalized states (as obtained for $\ket{v_n}$ uniformly spread over the computational basis), while $\mathrm{IPR} = O(1)$ if $\ket{v_n}$ is localized on a single Fock state. Figure~\ref{fig:vectors_scaled}a presents the logarithm of the IPR of singular vectors, scaled with system size: our data shows the presence of a finite-size crossover 
from delocalized to localized singular vectors
around a value $\gamma_c/J \approx 9$, consistent with the picture extracted from the average gap ratio ($r$-parameter) statistics.

We further present our results for the entanglement entropy across a bipartition in Fig.~\ref{fig:vectors_scaled}b. Recall that, for a generic state $\ket{\phi}$, the entanglement entropy is defined as $S_\mathrm{E} = - \tr \rho_A \ln \rho_A$, where $\rho_A = \Tr_B (\dyad{\phi}{\phi})$ and $A \cup B$ form a bipartition of the chain in two intervals.

As for the IPR, the entanglement entropy supports the presence of a localized regime: it crosses over from a volume law at small dissipation to an area law at large dissipation, again consistently indicating a critical value $\gamma_c/J\approx 9$ for the system sizes considered. Remarkably, the same analysis of the IPR and entanglement entropy with the eigenvectors does not display such a clear crossover~\cite{SupplMat}, thus strongly motivating the use of the SVD. While a weak, inhomogeneous dissipation breaks the integrability of the XXZ chain making it (dissipatively) chaotic, more disordered losses localize it again and restore integrability.


\textit{Conclusions.---}We have investigated the role of a disordered dissipative term on an otherwise integrable, interacting quantum system. Adding such a term makes the system evolve under an effective non-Hermitian (NH) Hamiltonian, physically representing the average evolution of quantum trajectories conditioned to no quantum jumps. The eigendecomposition of NH Hamiltonians yields complex eigenvalues and non-orthogonal (left and right) eigenvectors. We argued in favor of using the singular value decomposition and showed that, indeed, the singular values can be used to detect a crossover from chaotic to integrable spectral features, and the singular vectors can be used to probe a crossover from delocalization to localization. We introduced the singular form factor; it features a correlation hole when the dissipative disorder is weak, eventually closing for large dissipative disorder. In this setting, random dissipation-induced localization points to a quantum dynamics that is highly sensitive to the effect of inhomogeneities. This contrasts with a homogeneous dissipation which, in our case, does not induce localization. 

A crucial point, in the Hermitian setting, is how the chaotic/localized crossover scales with system size. This point has been highly debated in the last few years, and consensus has yet to be reached~\cite{Suntajs2020Quantum,Panda2020Can,Abanin2021Distinguishing,Sierant2022Challenges,Morningstar2022Avalanches,Crowley2022Constructive,Sierant2024Many}. For the dissipative case, this is not as relevant because the NH evolution describes an exponentially small (in system size and time) fraction of trajectories. In turn, our findings are meaningful especially for systems of small sizes: only in these cases can the chaotic/localized behaviors actually be observed in experiments. Our results are thus relevant for \textit{noisy intermediate-scale quantum} devices~\cite{Preskill2018Quantum}, since we show that the Hamiltonian properties are significantly altered by disordered dissipation.


\begin{acknowledgments}
\textit{Acknowledgments.---}We thank Masahito Ueda for useful discussions, and Oskar A. Pro\'sniak and Pablo Mart\'inez-Azcona for careful reading of the manuscript. F.B.~thanks Carlo Vanoni for illuminating discussions on localization indicators. This research was funded in part by the Luxembourg National Research Fund (FNR, Attract grant 15382998), the John Templeton Foundation (Grant 62171), and the QuantERA II Programme that has received funding from the European Union’s Horizon 2020 research and innovation programme (Grant 16434093). The numerical simulations presented in this work were partly carried out using the HPC facilities of the University of Luxembourg.

The opinions expressed in this publication are those of the authors and do not necessarily reflect the views of the John Templeton Foundation.
\end{acknowledgments}

\bibliographystyle{apsrev4-1}	
\bibliography{references}


\onecolumngrid
\appendix



\section{A. Effective Non-Hermitian Hamiltonians}
\label{App:effNHhamiltonian}

Several theoretical frameworks have been developed to describe the dynamics of open quantum systems: quantum trajectories~\cite{Brun2002Simple}, Zwanzig-Mori projection operators~\cite{ZwanzigPhysRev1961} and collision models~\cite{Ciccarello2022Quantum} to name a few. Here, we perform the common assumptions of having an infinitely large Markovian bath, with a fast relaxation timescale compared to that of the system, to which it is weakly coupled. Formally, this is reflected in the possibility of describing the dynamics with a Gorini–Kossakowski–Sudarshan–Lindblad (GKSL) master equation for the reduced density matrix $\rho_t$ of the system:
\begin{equation}
    \dot \rho_t = - i[\hat H,\rho_t] + 
    \sum_j 
    \mathcal D [\sqrt{\gamma_j} \hat L_j]\rho_t\,,
\end{equation}
with the dissipator  
$\mathcal{D} [\hat{\mathcal{O}}]\rho = \hat{\mathcal{O}}\rho \hat{\mathcal{O}}^\dagger - \tfrac{1}{2}\{\hat{\mathcal{O}}^\dagger \hat{\mathcal{O}},\rho\}$.
Above, the Hermitian Hamiltonian $\hat H$ describes the dynamics of the system alone, while the interaction with the bath is encoded in the jump operators $\hat{\mathcal{O}}= \sqrt{\gamma_j} \hat L_j$, $\gamma_j>0$. The introduction of an effective non-Hermitian (NH) Hamiltonian, 
\begin{equation}
    \label{eq:H_eff}
    \hat H_\text{eff} = \hat H - \frac{i}{2}\sum_j \gamma_j \hat L_j^\dagger \hat L_j,
\end{equation}
allows re-writing the GKSL equation as
\begin{equation}
    \dot \rho_t = - i (\hat H_\text{eff}\rho_t-\rho_t \hat H_\text{eff}^\dagger) + \sum_j \gamma_j \hat L_j\rho_t \hat L_j^\dagger.
\end{equation}
This equation separates the density matrix evolving under $\hat H_\text{eff}$ from the jump terms. 
The case we describe in the main text, Eq.~(1), is a NH Hamiltonian and corresponds to a dynamics with no jumps, governed by the effective Hamiltonian only. 

Two important remarks are in order. First, the use of the GKSL equation automatically implies that an average over the bath influence is taken \emph{at the level of the density matrix}. This fact implies that quantities that are non-linear in the density matrix, like entanglement, are not well captured within the GKSL framework~\cite{Li2018Quantum,Skinner2019Measurement}. Second, no-jump trajectories constitute an exponentially small fraction (both in time and in the system size) of all the quantum trajectories since, at each time (and for each spin, Eq.~(1) in the main text), one needs to impose that no jump occurs: these events are typically independent and their probabilities multiply. Despite these two caveats, we  nevertheless study a \emph{many-body} system under \emph{strong} local dissipation, and moreover focus on the MBL transition, which is usually linked to \emph{entanglement} properties. This might seem a contradiction, but it is only an apparent one. Indeed, the use of NH Hamiltonians has already been shown to capture well entanglement transitions in many-body quantum systems~\cite{SierantPRL2020,Biella2021Many,Turkeshi2021Measurement,Youenn2023Volume,TurkeshiPRB2023}. Additionally, averaging the density matrix \emph{before} computing the entanglement is reminiscent of the annealed average usually considered in disordered systems~\cite{Mezard1987Spin}, which provides an approximation that, in certain cases, is also accurate.


\section{B. General properties of the eigenvalue and singular value decomposition}
\label{App:EDvsSVD}

Here, we briefly review some known properties of the eigen- and singular-value decompositions of Hermitian and NH Hamiltonians. We interchangeably refer to the matrix representation of our Hamiltonian in some basis simply as the ``Hamiltonian'', as we focus on a finite-dimensional Hilbert space of dimension $D$. To fix the ideas, one can take the Hamiltonian $\hat H$ to describe the XXZ spin chain, Eq.~(2a) in the main text, in the zero magnetization sector, corresponding to $D = \binom{N}{N/2}$.

\subsection{Remarks on the eigendecomposition}

Consider a Hermitian Hamiltonian $H$ (we drop the hat as we refer to its matrix representation). It can be diagonalized as $H=W\Lambda W^\dagger=\sum_n E_n \dyad{w_n}$ through a single unitary transformation, $W=(\ket{w_1},\ldots,\ket{w_\textsc{d}})$,  $\Lambda=\diag(E_1,\ldots,E_\textsc{d})$ being the diagonal matrix formed by the real eigenvalues. The eigenstates of $H$, $\{\ket{w_n}\}$, are orthogonal and can be made orthonormal; so, they resolve the identity,  $\mathbb{1}_\textsc{d}=\sum_n \dyad{w_n}$. The eigenvalues and eigenstates of a Hermitian operator $H$ have a physical meaning in quantum mechanics, being respectively the outcomes of the measurements of $H$ and the corresponding states the system collapses to after the measurement.

In turn, a non-Hermitian Hamiltonian $H\neq H^\dagger$ is not guaranteed to be diagonalizable. The lack of diagonalizability comes from a degeneracy of eigenvalues \textit{and} eigenvectors, at what are known as \textit{exceptional points}~\cite{Miri2019Exceptional}. Away from these degeneracies, we can assume $H$ to be diagonalizable, although its eigenvectors are in general not orthogonal and its eigenvalues can be complex. A way out of this, that has been intensely used in the NH physics literature~\cite{Brody2014Biorthogonal}, is to use both the right and left eigenvectors of $H$, satisfying $H\ket{R_n}= E_n\ket{R_n}$ and  $\bra{L_n} H= E_n\bra{L_n}$, respectively. These are not orthonormal sets themselves, as $\braket{R_n}{R_m}\neq \delta_{nm}$ (same with $L$), but are \textit{biorthogonal}, \textit{i.e.}, $\braket{L_n}{R_m}= c\, \delta_{nm}$. The constant $c$ can be set to one to make the right and left eigenvectors \textit{biorthonormal}. This way, the identity can be resolved as  $\mathbb{1}_\textsc{d}=\sum_n \dyad{R_n}{L_n}$. Note that, if the right and left eigenvectors are biorthonormal, one can still normalize the right ones but not the left ones, and vice versa. Assuming biorthonormality, one can make the decomposition $H=R\Lambda L^\dagger=\sum_n E_n\dyad{R_n}{L_n}$, where $R=(\ket{R_1},\ldots,\ket{R_\textsc{d}})$ and $L=(\ket{L_1},\ldots,\ket{L_\textsc{d}})$ are the non-unitary transformations that make $H$ diagonal, and where the eigenvalues $E_n$ are now generally complex.

As we mention in the main text, there is a sort of ambiguity in the literature on how to properly define the density matrix corresponding, \textit{e.g.}, to the right eigenvector $\ket{R_n}$. On the one hand, if one wants to keep the expectation values of an observable $\hat O$ real (as much as possible~\cite{Brody2014Biorthogonal}), the expectation value of $\hat O$ on the right eigenvector $\ket{R_n}$ must be defined as $\langle\hat O\rangle=\bra{L_n}\hat O\ket{R_n}$ (assuming $\braket{L_n}{R_m}= \delta_{nm}$). This leads to the conclusion that the density matrix corresponding to $\ket{R_n}$ should be $\rho_\text{RL}=\dyad{R_n}{L_n}$, so that $\langle\hat O\rangle=\text{tr}(\hat O \rho_\text{RL})$ where $\text{tr}(\ldots)=\sum_n\bra{L_n}\ldots\ket{R_n}$ ~\cite{Brody2014Biorthogonal}. Though preserving the probabilistic interpretation, such a density matrix is no longer Hermitian. On the other hand, in order to preserve Hermiticity, the entanglement entropy of a right eigenvector $\ket{R_n}$ is typically defined using  $\rho_\text{RR}=\dyad{R_n}{R_n}$ as a definition of the corresponding density matrix~\cite{Hamazaki2019NonHermitian}.

\subsection{Brief review of the singular value decomposition}

Any Hamiltonian $H$, be it Hermitian or not, can be decomposed through the singular value decomposition (SVD) into $H=U\Sigma V^\dagger = \sum_n \sigma_n \dyad{u_n}{v_n}$, where $U=(\ket{u_1},\ldots,\ket{u_\textsc{d}})$ and $V=(\ket{v_1},\ldots,\ket{v_\textsc{d}})$ are unitaries whose columns are the left and right \textit{singular vectors}, respectively, and $\Sigma=\diag(\sigma_1,\ldots,\sigma_\textsc{d})$ is the positive semi-definite matrix of \textit{singular values}. 
As $U$ and $V$ are unitaries, their columns are orthonormal and resolve the identity without any need to recur to bi-orthogonality.

The SVD makes explicit how a general matrix $H$ acts on an orthonormal basis $V$, to get a rotated orthonormal basis $U$, stretched or compressed by the real and non-negative numbers in the diagonal of $\Sigma$.
From an operational point of view, one can obtain the SVD of a matrix by using an eigendecomposition procedure. The basis $V$ is the orthonormal eigenbasis of the Hermitian matrix $H^\dagger H$ with eigenvalues $\{\sigma_n^2\}$, while the orthonormal basis $U$ is the eigenbasis of the Hermitian matrix $HH^\dagger$, sharing the same eigenvalues. By standard manipulations in linear algebra, one can see that the SVD of a matrix $H$ can also be obtained from the eigendecomposition of the \emph{cyclic matrix}
\begin{equation}
    \label{eq:C}
    C =
    \begin{pmatrix}
        \mathbb{0}  &H  \\
        H^\dagger   &\mathbb{0}
    \end{pmatrix}.
\end{equation}
$C$ is a $2D \times 2D$ matrix with eigenvalues $\{\pm \sigma_n\}$, associated to eigenvectors $(\pm u_n, v_n)^T / \sqrt{2}$.

It is important to stress that the two procedures described above to obtain the SVD of a matrix from eigendecomposition routines are equivalent in principle, but different in practice. The use of either $H^\dagger H$ or $H H^\dagger$ does not increase the matrix dimension, but yields the singular values squared. This results in a loss of precision, which is particularly severe for the small singular values---which are the ones that we used in the main text. On the other hand, the cyclic matrix $C$ provides directly the singular values $\sigma$, albeit with a doubled computational cost (typically, one gets both $\sigma_n$ and $-\sigma_n$), and with the caveat that the small $\sigma_n$'s lie at the center of the spectrum of $C$---and procedures as shift-invert or polynomial filtering must be used. 

Finally, beyond the reasons mentioned in the main text, there is yet another motivation to use the SVD for NH Hamiltonians. For Hermitian Hamiltonians, a constant diagonal shift \textit{does not} affect the physics; the ED captures this fact, 
as the eigenvalues are rigidly shifted and the eigenstates unaffected. However, the SVD is  sensitive to a diagonal shift, making its physical meaning rather questionable in  Hermitian systems. By contrast, any diagonal shift of a NH Hamiltonian \textit{does change} the physics, as exemplified in a vectorized Lindbladian (a NH matrix) that admits a steady state (zero eigenvalue): a rigid spectral shift along the real axis would either introduce amplified modes or increase the decay rates, while a shift along the imaginary axis could remove the steady state. Since such a shift is captured by the SVD, it seems a more suited decomposition to study NH Hamiltonians.

\section{C. Complex spectral gap ratios} 
\label{App:complex_r_ratios}

In the case of an effective open system Hamiltonian such as Eq.~\eqref{eq:H_eff} the spectrum is complex and there is no clear notion of ordering the eigenvalues. One can nevertheless find the nearest neighbor $E_n^{\textsc{nn}}$ of each eigenvalue $E_n$ on the complex plane, defined as the eigenvalue $E_n^{\textsc{nn}}$ which satisfies $ |E_n^{\textsc{nn}}-E_n|=\min_{m\neq n} |E_n-E_m|$.
As explained in the main text, to avoid the need to unfold the spectrum, we would like to consider a ratio of energy-dependent quantities. A natural other such quantity would be the next-nearest neighbor $E_n^{\textsc{nnn}}$ for each eigenvalue, which would amount to ordering the set $\{|E_m-E_n|\}_{m\neq n}$ and selecting the second entry to find $E_n^{\textsc{nnn}}$. The spectral ratio to be studied is then the complex ratio~\cite{Sa2020Complex}
\begin{equation} \label{complex_ratio}
    z_n \equiv \frac{E_n^{\textsc{nn}}-E_n}{E_n^{\textsc{nnn}}-E_n} ~.
\end{equation}
Note that this definition is similar but not the same as the ratio of closest-neighbor and 2nd closest-neighbor defined in Ref.~\cite{Srivastava2018Ordered} for real eigenvalues. In that work, the ratio of the absolute value of the difference between levels (\textit{i.e.}, distance) was considered, while in the definition~\eqref{complex_ratio} it is not the ratio of distances which is computed, but rather the ratio of differences.  More specifically, the ratio for real eigenvalues defined in Ref.~\cite{Srivastava2018Ordered} can take values between 0 and 1, while the definition~\eqref{complex_ratio}, restricted to real eigenvalues, gives values between $-1$ and $1$.

As in the Hermitian case, it is often useful to extract a single number from the ratio statistics as a probe to the type of spectrum. In the complex case $z_n = r_n e^{i \theta_n}$ with $0\leq r_n \leq 1$ because $|E_n^{\textsc{nn}}-E_n| \leq |E_n^{\textsc{nnn}}-E_n|$. One such single-number probe would be the averages $r = \langle r_n \rangle$ and $\cos \theta = \langle \cos \theta_n \rangle$.
Table~\ref{tab:CR_table} shows the values expected from uncorrelated complex eigenvalues (Poisson ensemble) and random matrix ensembles, as reported in Ref.~\cite{Sa2020Complex}. The Ginibre Unitary Ensemble (GinUE) are complex matrices with independent identically distributed complex entries taken from a Gaussian distribution and setting $\beta=2$ in the joint eigenvalue distribution. The Toric Unitary Ensemble (TUE) is a generalization of the Circular Unitary Ensemble (where eigenvalues lay on a circle) to the complex plane compactified into a torus (such that all eigenvalues lay on the torus). The next Section discusses this single-number probe in our model. 

\begin{table}
    \centering
    \begin{tabular}{|c|c|c|c|}
        \hline
         &  Poisson & GinUE & TUE\\ 
         \hline
        $ r $  & 2/3 &   0.73810 & 0.7315 -- 0.73491\\
        $- \cos \theta $ & 0 &  0.24051 &  0.15322 --  0.1938 \\
        \hline
    \end{tabular}
    \caption{Average values expected for $\langle r_n \rangle$ and $-\langle \cos \theta_n \rangle$ for different types of complex eigenvalue statistics, taken from Ref.~\cite{Sa2020Complex}.}
    \label{tab:CR_table}
\end{table}

\begin{figure*}
    \includegraphics[width=0.7\textwidth]{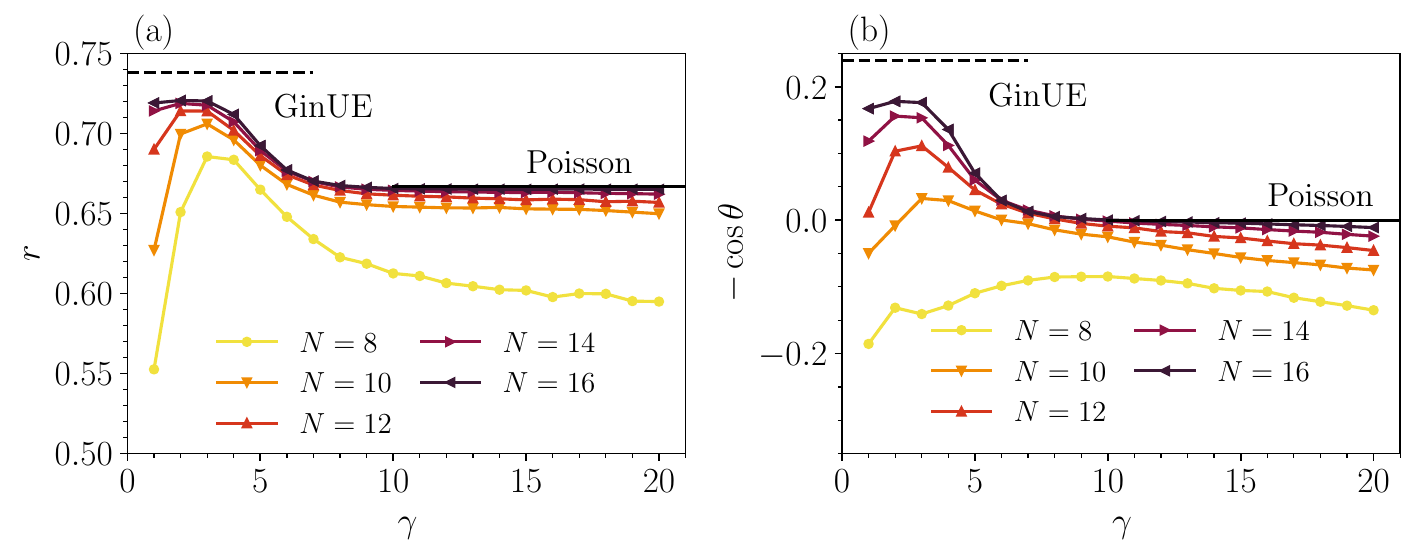}
    \caption{\textit{Eigenvalue statistics for the XXZ model with random losses using complex gap ratio statistics.} (a) The average value of the norm of the complex gap ratios. (b) Cosine of the phase of the complex gap ratios. The data was extracted from a small window in the middle of the (complex) spectrum and averaged over at least 7000 disorder realizations. Here and in all successive figures, error bars are smaller than the displayed dots.}
    \label{fig:EIGvalues}
\end{figure*}

\section{D. Further results for the XXZ model with random losses}
\label{App:furtherXXZ}

In the main text, we studied the \textit{real} singular-value statistics for the model~(1) and found they exhibit a clear crossover from GOE to Poisson, see Fig.~1. Let us now present the eigenvalue statistics using the \textit{complex} spectral gap ratios as described in Section~\ref{App:complex_r_ratios}. Figure~\ref{fig:EIGvalues} displays the gap ratio statistics for the complex eigenvalues of the XXZ model with dissipative disorder, Eq.~(1) in the main text, as a function of the disorder strength $\gamma$. As we show in Fig.~\ref{fig:EIGvalues}, the results for $r$ and $\cos \theta$ do not provide clear-cut values which we could clearly attribute to a crossover from chaos to integrability. Also, the behavior of these single-number statistics does not change smoothly with the system size $N$, in contrast with the behavior of the single-number statistics for the singular values.

\begin{figure*}
    \includegraphics[width=\textwidth]{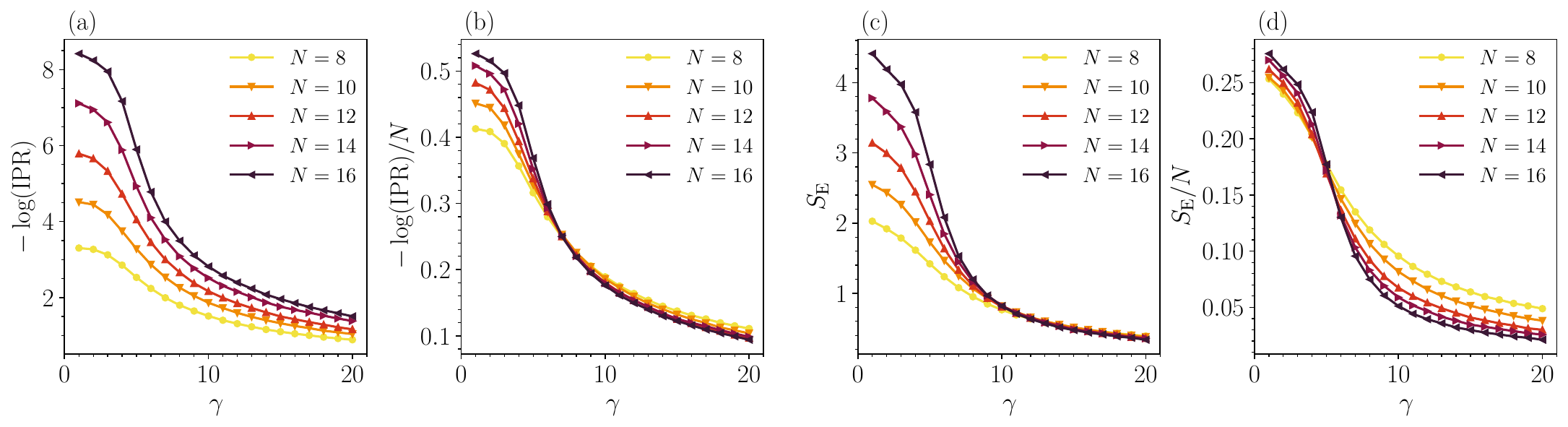}
    \caption{\textit{Localization indicators for the right eigenvectors of the  XXZ model with random losses.} 
    (a) and (c): IPR and entanglement entropy of the right eigenvectors the  XXZ model with random losses, respectively. (b) and (d), same plots including a scaling with system size. Both indicators show a localization transition compatible with the one observed for singular vectors. The data was extracted from a small window in the middle of the (complex) spectrum and averaged over at least 7000 disorder realizations.}
    \label{fig:EIGvectors}
\end{figure*}

We present as well a study on localization indicators for the (right) eigenvectors of the NH XXZ model, Eq.~(1) main text, in Fig.~\ref{fig:EIGvectors}. For the entanglement entropy, we use the definition $S_\mathrm{E} = - \tr \rho_A \ln \rho_A$ with $\rho_A = \Tr_B (\dyad{R_n}{R_n})$, \emph{i.e.},\ using the right eigenvectors only~\cite{Hamazaki2019NonHermitian}. A crossover between a delocalized/localized regime for the eigenvectors occurs similarly to the one observed for the singular vectors. In principle, however, both right and left eigenvectors could be used to construct the reduced density matrix~\cite{lu2023unconventional}, even if the physical difference between the various cases is not yet well understood. By contrast, the singular vectors constitute an orthonormal basis and the definition of the density matrix is more direct. This supports even more the use of the SVD as a tool to generalize standard Hermitian phenomena, as already motivated in totally different settings~\cite{HerviouPRA2019,BrunelliSciPost2023}.

\begin{figure*}
    \includegraphics[width=\textwidth]{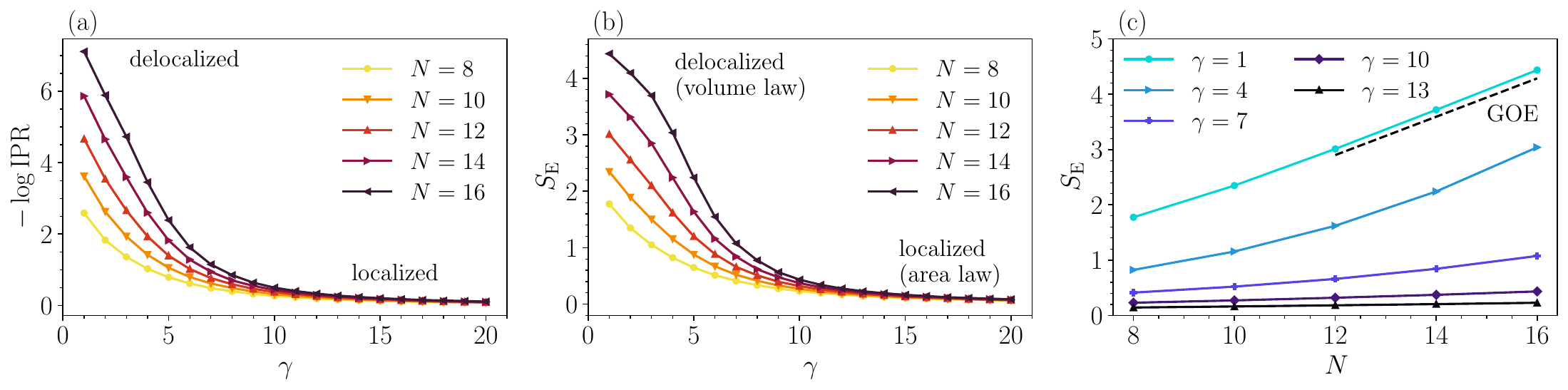}
    \caption{\textit{Dissipative localization of singular vectors (without rescaling) of the XXZ model with random losses.} Here we plot the same data shown in Fig.~2 in the main text without rescaling by system size. (a) The inverse participation ratio (IPR) quantifies how much a state is localized in the many-body Fock space, passing from $\mathrm{IPR} \simeq 1/D$ (delocalized) for weak dissipation to $\mathrm{IPR} \simeq 1$ (localized) for strong dissipation. Our data shows a crossover between the two regimes as the disorder in the dissipator is ramped up. (b) Entanglement entropy across a bipartition of the chain in two halves. A delocalized regime is visible for small dissipation $\gamma$, while localization takes place at larger $\gamma$. (c) The entanglement entropy changes from a volume law $S_\mathrm{E} \sim N \ln(2)/2$ (small $\gamma$), as expected for a state completely delocalized in the Fock basis (dashed black line), to an area law $S_\mathrm{E} \sim \mathrm{const.}$ (large $\gamma$), typical of localized states. All the data was extracted from a small window in the middle of the spectrum and averaged over at least 7000 disorder realizations.}
    \label{fig:vectors}
\end{figure*}

Finally, in the main text, we showed the localization crossover captured by the scaled IPR and entanglement entropy. Here, we further show the same results, without the system-size rescaling in Fig.~\ref{fig:vectors}a--b, which masks the crossover, but clearly show the transition between volume law to area law, Fig.~\ref{fig:vectors}c.

\section{E. Results for the interacting Hatano-Nelson model}
\label{App:resultsHatano}

Here we perform the same analysis described in the main text for the interacting Hatano-Nelson model
\begin{equation}
    \label{eq:Hatano-Nelson-Model}
    \hat H  = J\sum_{i=1}^N \left[ \frac{1}{2} \left( e^g \hat S_i^+ \hat S_{i+1}^- + e^{-g} \hat S_i^- \hat S_{i+1}^+ \right) + \Delta \hat S_i^z \hat S_{i+1}^z \right] + \sum_{i=1}^N h_i \hat S_i^z .
\end{equation}
This model was studied in Ref.~\cite{Hamazaki2019NonHermitian}, where it was shown it hosts a NH MBL phase. Here, we study its localization properties through the lens of the SVD. We fix, as for the model in the main text, $J=1$ and $\Delta = 1$, and further set $g=0.1$. Then, we extract the disordered fields $h_i \in [-h,h]$ according to the uniform distribution. In order to avoid the NH skin effect, which may be mistaken for localization, we use periodic boundary conditions.

In contrast with the Hamiltonian considered in Eq.~(1) in the main text,  where non-Hermiticity and disorder coincide, the non-Hermiticity in the interacting Hatano-Nelson model, Eq.~\eqref{eq:Hatano-Nelson-Model}, comes from the unbalanced hoppings while the randomness comes from a standard (Hermitian) local field. Therefore, in this model, localization and chaos are not \emph{induced} by non-Hermiticity, but one can quantify their \emph{resilience} to non-Hermiticity.  

\begin{figure*}
    \includegraphics[width=\textwidth]{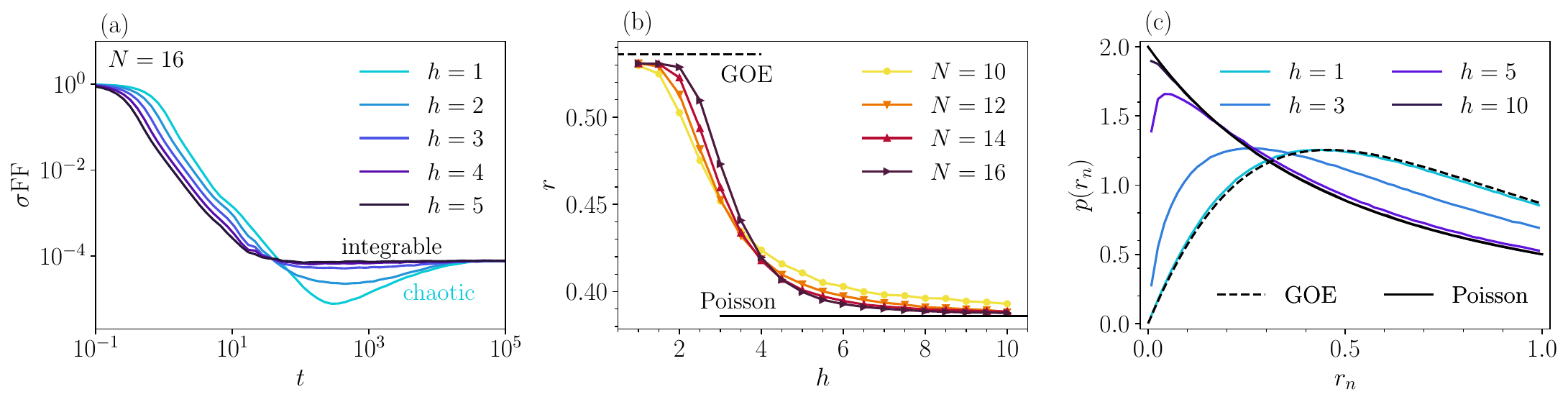}
    \caption{\textit{Singular value statistics for the interacting Hatano-Nelson model with $g=0.1$ and $J=1=\Delta$.} (a) The $\sigma$FF shows that as $h$ increases correlations among the singular values disappear and the correlation hole closes. (b) The average of the ratio distributions, for data taken from the smallest singular values, shows a clear crossover from the GOE (chaotic) value to the Poisson (integrable) value as $h$ is increased, with the transition around $h/J\approx 4$. (c) The ratio statistics follow the distribution expected from the GOE for a small value of $h$, and that expected from the Poissonian ensemble for a large value of $h$ (shown for $N=16$).  All data is averaged over at least 5000 realizations of the disorder.}
    \label{fig:HNvalues}
\end{figure*}

\begin{figure*}
    \includegraphics[width=\textwidth]{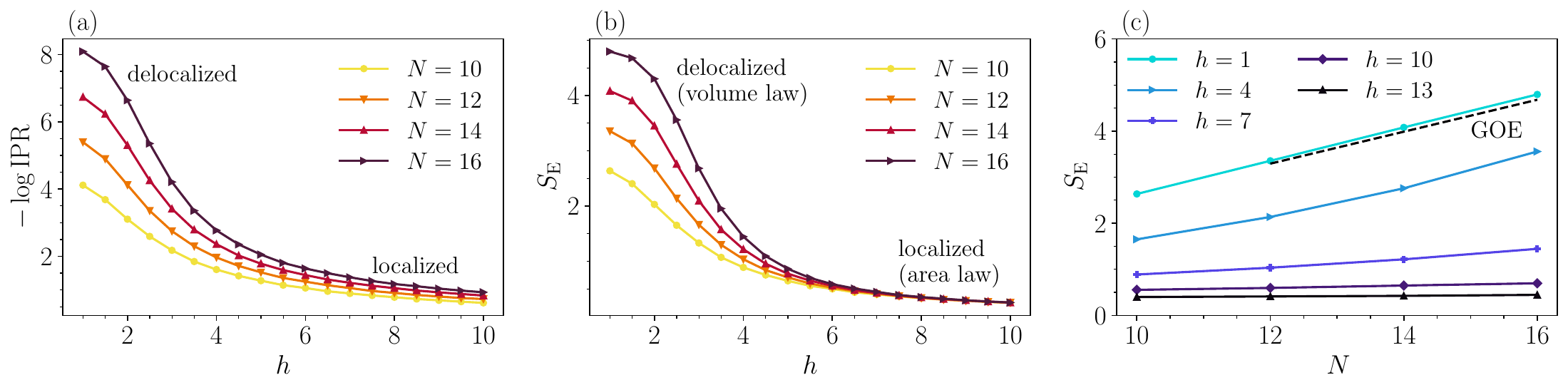}
    \caption{\textit{Dissipative localization of singular vectors for the interacting Hatano-Nelson model.} (a) The IPR as a function of $h$, plotted for several system sizes $N$. For all $N$'s, the IPR shows a crossover from delocalized singular vectors to localized singular vectors as $h$ increases. (b) The entanglement entropy of the singular vectors as a function of $h$, for the same system sizes. Also $S_E$ shows a clear crossover between delocalization (volume law) and localization (area law) as $h$ is increased. (c) The crossover from $S_\mathrm{E} \sim N \ln(2)/2$ (volume law, dashed black line) to $S_E \sim \mathrm{const.}$ (area law) as $h$ is increased, now plotted as a function of the system size $N$. All data is averaged over at least 5000 realizations of the disorder.}
    \label{fig:HNvectors}
\end{figure*}

\begin{figure*}
    \includegraphics[width=0.7\textwidth]{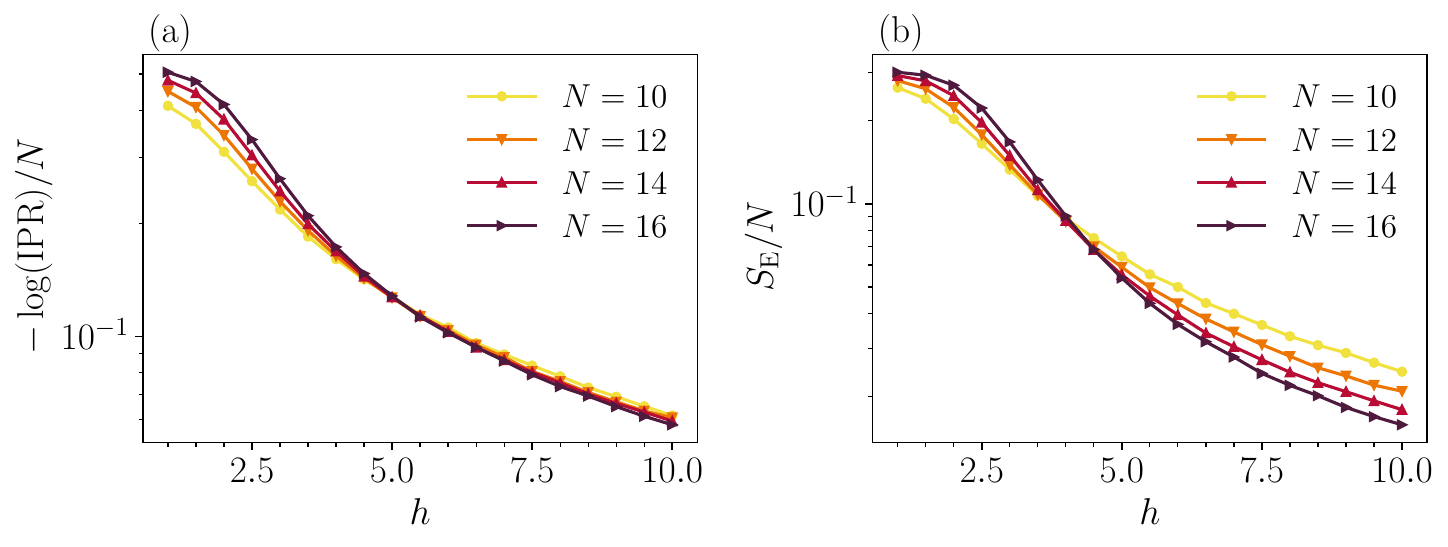}
    \caption{\textit{Dissipative localization of singular vectors for the interacting Hatano-Nelson model (with rescaling).} Same data as in Fig.~\ref{fig:HNvectors}, where both (a) the IPR and (b) the entanglement entropy are rescaled with system size to highlight the crossover.}
    \label{fig:HNvectors_scaled}
\end{figure*}

The singular form factor and the singular value statistics for the interacting Hatano-Nelson model, Eq.~\eqref{eq:Hatano-Nelson-Model}, as function of disorder strength $h$, are shown in Fig.~\ref{fig:HNvalues}. The localization properties of the singular vectors, as quantified by the IPR and the entanglement entropy, are plotted as a function of $h$ in Figs.~\ref{fig:HNvectors}--\ref{fig:HNvectors_scaled}.

\end{document}